\documentclass[]{spie}  %>>> use this instead for A4 paper
\usepackage{amsmath,amsfonts,amssymb}
\usepackage{graphicx}
\usepackage[colorlinks=true, allcolors=blue]{hyperref}

\usepackage{subfigure}
\graphicspath{{Figures/}}
\usepackage{algorithm}
\usepackage{algorithmicx}
\usepackage{algpseudocode}
\usepackage{bm}
\usepackage{textcomp}
\usepackage[dvipsnames]{xcolor}
\usepackage{multirow}
\def\eg{{\it e.g., }}

\newcommand{\etc}{{etc.}}              
\def\be{\begin{equation}}
\def\ee{\end{equation}}
\def\ba{\begin{array}}
\def\ea{\end{array}}
\def\ban{\begin{eqnarray}}
\def\ean{\end{eqnarray}}
\everymath{\displaystyle}

\usepackage[utf8]{inputenc}
\usepackage[english]{babel}
\usepackage{hyperref}

\title{Lesion detection in Contrast Enhanced Spectral Mammography}

\author[]{Clément Jailin}
\author[]{Pablo Milioni}
\author[]{Zhijin Li}
\author[]{Răzvan Iordache}
\author[]{Serge Muller}

\affil[]{GE Healthcare, 78530 Buc, France}

\authorinfo{C.J.: E-mail: clement.jailin@ge.com}

\begin{document} 
\maketitle

\begin{abstract}
\paragraph{Background \& purpose:} The recent emergence of neural networks models for the analysis of breast images has been a breakthrough in computer aided diagnostic. This approach was not yet developed in Contrast Enhanced Spectral Mammography (CESM) where access to large databases is complex.
This work proposes a deep-learning-based Computer Aided Diagnostic development for CESM recombined  images able to detect lesions and classify cases.
\paragraph{Material \& methods:}  A large CESM diagnostic dataset with biopsy-proven lesions was collected from various hospitals and different acquisition systems. The annotated data were split on a patient level for the training (55\%), validation (15\%) and test (30\%)  of a deep neural network with a state-of-the-art detection architecture. Free Receiver Operating Characteristic (FROC) was used to evaluate the model for the detection of 1) all lesions, 2) biopsied lesions and 3) malignant lesions. ROC curve was used to evaluate breast cancer classification. The metrics were finally compared to clinical results.
\paragraph{Results:}  For the evaluation of the malignant lesion detection, at high sensitivity (Se$>$0.95), the false positive rate was at 0.61 per image. For the classification of malignant cases, the model reached an Area Under the Curve (AUC) in the range of clinical CESM diagnostic results.
\paragraph{Conclusion:} This CAD is the first development of a lesion detection and classification model for CESM images. Trained on a large dataset, it has the potential to be used for helping the management of biopsy decision and for helping the radiologist detecting complex lesions that could modify the clinical treatment.
 
\end{abstract}

% Include a list of keywords after the abstract 

% =========================================================
\section{INTRODUCTION}
% =========================================================
 
The recent development of Computer Aided Detection (CADe) and Computer Aided Diagnostic (CADx) based on Deep Learning (DL) has been a breakthrough in medical imaging~\cite{zhou2017deep,sahiner2019deep,fujita2020ai} and in breast cancer analysis. 
In breast images, various CAD exist for screening or diagnostic applications and are used on Full Field Digital Mammography~\cite{shen2019deep,kim2020changes} or Digital Breast Tomosynthesis\cite{al2018simultaneous} where large databases may be available for the training. 

Contrast-enhanced spectral mammography (CESM) provides anatomical and functional imaging of breast tissue improving the accuracy of breast cancer diagnosis~\cite{dromain2011dual,luczynska2014contrast}. This recent imaging technique has started to attract broad research and clinical application interest.~\cite{diekmann2005digital,puong2007dual,lewin2020contrast}. After an intravenous iodine injection, the compressed breast is imaged with two X-ray energies. The iodine-equivalent image (called recombined image) is obtained from processing of the low and high energy images. 
%CESM sensitivity in breast cancer detection is similar to breast MRI imaging techniques with significantly shorter exam time and lower costs~\cite{fallenberg2014contrast,fallenberg2017contrast}.

In Contrast Enhanced Digital Mammography application, the current developments in artificial intelligence are focused on lesion classification to predict the pathology results. Due to the small size of available databases (generally less than 130 patients), the approaches are often based on the classification of manually extracted regions of interests. From those images, handcrafted features (\eg radiomics\cite{losurdo2019radiomics,patel2018computer,lin2020contrast,massafra2021radiomic}
) or deep learning features (\eg obtained from a convolutional neural network \cite{perek2019classification}) are extracted and fed into machine learning algorithms (\eg support vector machine, multi-layer perceptron, \etc) to perform the classification.
Those analyses require the suspicious area to be detected and extracted manually by the user. In some applications \cite{perek2019classification,caballo2021multi}, a pixel segmentation of the contrast uptakes may be necessary.

The current study is a first development of an automatic lesion detection and analysis model in CESM images. A deep learning model is trained on a large database (586 patients) of diagnostic CESM patients to localize contrast uptakes in the iodine images and associates them a suspicion score that would aid the clinician in their diagnostic. The results are finally compared with clinical diagnostic results.

% =========================================================
\section{Data and Method}
% =========================================================

\paragraph{CESM datasets} A CESM dataset consisting of 586 patients with 2510 recombined images of left/right views, mainly cranio-caudal (CC) and mediolateral oblique (MLO), was collected from various hospitals and acquired with different systems (Senographe DS, Essential and Pristina from \emph{GE Healthcare, Chicago, Illinois, United States}). The lesions in the dataset were all biopsy proven. The number of normal, benign and malignant cases are respectively 191 (33\%), 149 (25\%) and 246 (42\%). When training a detection model, having normal cases allows reducing the false positive detection by training to recognize what is not a lesion. It is expected that a too large proportion of normal cases may also degrade the model efficiency. The optimal quantity of normal / annotated data was not studied in this paper.
%\SM{Exp:liquer pourquoi et commenter en quoi la proportion de cas normaux dans la database est un avantage ou une limitation.}. 
Note that for the cases where a lesion is detected, the contralateral breast may be labeled normal if no findings are detected. It therefore  represents normal cases for the training. 
The summary of the data is provided in table~\ref{tab:1}.

\begin{table}[ht]
\centering
\begin{tabular}{|l|l|l|lll|}
\hline
\multicolumn{1}{|c|}{\multirow{2}{*}{\textbf{Hospital}}} &
  \multicolumn{1}{c|}{\multirow{2}{*}{\textbf{Num of patients}}} &
  \multicolumn{1}{c|}{\multirow{2}{*}{\textbf{Acq. sys.}}} &
  \multicolumn{3}{c|}{\textbf{Pathology results}} \\ \cline{4-6} 
\multicolumn{1}{|c|}{} &
  \multicolumn{1}{c|}{} &
  \multicolumn{1}{c|}{} &
  \multicolumn{1}{c|}{Normal} &
  \multicolumn{1}{c|}{Benign} &
  \multicolumn{1}{c|}{Malignant} \\ \hline
\begin{tabular}[c]{@{}l@{}}Peking University \\ First Hospital \&\\ Shanghai First  People's \\ Hospital (China)\end{tabular} &
  244 (976 images) &
  DS / Essential &
  \multicolumn{1}{l|}{35 (14\%)} &
  \multicolumn{1}{l|}{64 (26\%)} &
  145 (60\%) \\ \hline
  \begin{tabular}[c]{@{}l@{}}CBIS - Carolina Breast \\ Imaging Specialists (US)\end{tabular}&
  26 (143 images) &
  Pristina &
  \multicolumn{1}{l|}{1 (5\%)} &
  \multicolumn{1}{l|}{5 (19\%)} &
  20 (77\%) \\ \hline
   \begin{tabular}[c]{@{}l@{}}Beth Israel Deaconess \\ Medical Center (US)\end{tabular}&
  50 (211 images) &
  Essential &
  \multicolumn{1}{l|}{2 (4\%)} &
  \multicolumn{1}{l|}{24 (48\%)} &
  24 (48\%) \\ \hline
  \begin{tabular}[c]{@{}l@{}}University of \\ Cambridge (UK)\end{tabular}&
  39 (156 images) &
  Pristina &
  \multicolumn{1}{l|}{0 (0\%)} &
  \multicolumn{1}{l|}{11 (28\%)} &
  28 (72\%) \\ \hline
\begin{tabular}[c]{@{}l@{}}University of \\ Washington (US)\end{tabular} &
  187 (790 images) &
  Essential &
  \multicolumn{1}{l|}{153 (81\%)} &
  \multicolumn{1}{l|}{30 (2\%)} &
  4 (81\%) \\ \hline
\begin{tabular}[c]{@{}l@{}}Hospital del Mar (Spain)\end{tabular} &
  40 (234 images) &
  Pristina &
  \multicolumn{1}{l|}{0 (0\%)} &
  \multicolumn{1}{l|}{15 (38\%)} &
  25 (62\%) \\ \hline 
\end{tabular}
\caption{Details on the CESM dataset and pathology results.}
\label{tab:1}
\end{table}

For the training of the deep learning models, the 586 patients data were split in a training (332 patients - 1501 images) / validation (82 patients - 332 images) / test sets (172 patients - 677 images) stratified by the pathology results (normal/benign/malignant) and by clinical sites. 
In addition to the biopsied lesions, all other benign lesions in the breast (\eg non-biopsied cysts, fibroadenoma, nodes, \etc) were annotated with a rectangular bounding box and labeled with a specific label.

All CESM recombined data were processed using the latest \emph{GE Healthcare} commercial CESM recombination algorithm (Nira processing) to reduce artifacts. 
To be used in the deep learning framework, the images were pre-processed.  The image intensities were bounded between 1950 and 2205, corresponding to most of the recombined breast and iodine intensity range and converted into 8 bits.

\paragraph{Method:} The deep learning detection model used in this study is a small Yolo-v5s model. This is a PyTorch implementation of the similar Yolo-v4~\cite{bochkovskiy2020yolov4}, a state-of-the-art model for object detection. With a scalable architecture, it can be trained with relatively small datasets. From the published Yolo-v4, the only evolution of v5 used in this study is the automatic anchor size estimation based on k-means.

The architecture consists in 3 parts: (a) a CSPDarknet backbone to extract features at different scales, (b) a PANet neck to perform features combination using a feature pyramidal structure and (c) a Yolo Layer head that generates feature maps at three different scales allowing to identify small, intermediate and large lesions. The multi-scale characteristics of this model ensures a correct determination of all the lesion sizes (from a diameter $<3$~mm up to $>10$~cm).

We used as loss function $\mathcal{L}_{\text{DIoU}}$ a Distance Intersection over Union (DIoU) loss~\cite{zheng2020distance} that simultaneously considers the overlapping area and the distance to the center of the ground truth boxes. The IoU is defined as the ratio between the area of overlap over the area of union between the detected and ground truth bounding boxes. 
\begin{equation}
    \mathcal{L}_{\text{DIoU}}=1-\text{IoU}+\dfrac{\rho^2(\bm x, \bm x^{gt})}{c^2}
\end{equation}
with $\rho$ the Euclidean distance, $\bm x$ and $\bm x^{gt}$ respectively the coordinates of the centers for the predicted and ground truth boxes and $c$ the diagonal length of a rectangle circumscribed to the predicted and ground truth boxes. This loss leads to a better convergence than simple IoU loss. It can be noted that other metrics (\eg IoU, Complete IoU, Generalized IoU) were tested yet not providing better results than DIoU. Complete IoU (CIoU), considering the aspect ratio of the bounding boxes, gave similar results certainly because the shapes of the annotations are not very discriminant in lesion detection applications.

To compensate for small (compared to Deep learning standards) datasets, two techniques were applied: data-augmentation strategies and transfer learning. Data-augmentation strategies suited to breast images were used such as image flips, global intensity transforms and dedicated breast geometrical realistic transforms as developed in a previous study~\cite{caselles2021data}. The weights of the model were initially pre-trained on Image-Net, then re-trained on the lesion detection. Finally, the inference is performed with Test Time Augmentation (TTA) that consists in performing the inference on an image duplicated with simple transforms (scales and flips) and averaging all detections. 
After the inference, a non-maximum suppression (NMS) with a IoU threshold of 0.2 is used to reduce the number of superimposed detections.

The optimization was performed with a stochastic gradient descent ($lr=0.01$, momentum$=0.937$). The batch size was set to 12 images to fit the GPU memory (24Gb - Quadro RTX-6000). Approximately 500 iterations were necessary to reach convergence.

To validate the detection of an ROI, an acceptation criterion based on the IoU with the ground truth is used (threshold sets to IoU$_{\text{threshold}}=0.3$). 
The results for the lesion detection are evaluated with FROC (Free Receiver Operating Characteristic) curves considering 3 metrics:
\begin{enumerate}
    \item the sensitivity for the detection of all lesions (biopsied and not) in the breast
    \item the sensitivity for the detection of only biopsied lesions (identified suspicious by the radiologist). A clinical use of this metric would be suspicious lesion detection in screening CESM. This application may help the radiologist for the analysis of complex cases (\eg small and blurry lesion, satellite lesion).
    \item the sensitivity for the detection of cancers. This metric corresponds to a CAD aiming to reduce negative biopsies.
\end{enumerate}
In addition, the predicted suspicion score associated to each detected lesion allows performing a more general classification at image, breast or patient level. Each breast suspicion score is hence defined by the score of the most suspicious detected lesion. This classification will be evaluated in term of sensitivity and specificity with a ROC curve and compared with results from radiologists extracted from a CESM diagnostic clinical review paper~\cite{zhu2018diagnostic}.

% =========================================================
\section{Results}
% =========================================================

\begin{figure}[!ht]
	\centering
        \subfigure[]{\includegraphics[width=0.45\textwidth]{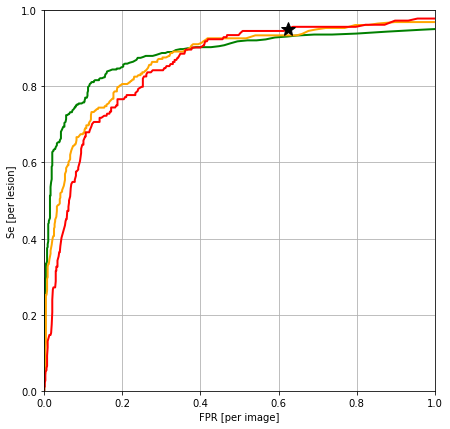}}
        \subfigure[]{\includegraphics[width=0.45\textwidth]{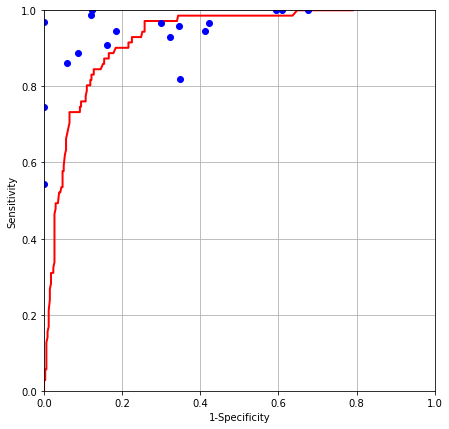}}
	\caption{
	(a) FROC curve with the metrics defined in section 2 - green: 1, - orange: 2 and - red: 3. A marker is plotted for a sensitivity of 0.95 (star).
	(b) ROC curve for cancer classification at the breast level (blue dots correspond to radiologists' diagnostic results from~\cite{zhu2018diagnostic}).}
	\label{fig:1} 
\end{figure}

The different FROC curves corresponding to the 3 defined metrics are plotted in figure~\ref{fig:1}(a). 
With a score threshold at 0.1 corresponding to a sensitivity of 0.95 for the detection of cancers (corresponding to the metric 3.), the false positive rate is 0.61 FP per breast. This metric is presented by the red FROC curve. The maximal detection sensitivity for cancers is 0.989 (with all detection scores accepted) with 3.8 FP per breast. 

\begin{figure}[!htbp]
	\centering
        \includegraphics[width=0.8\textwidth]{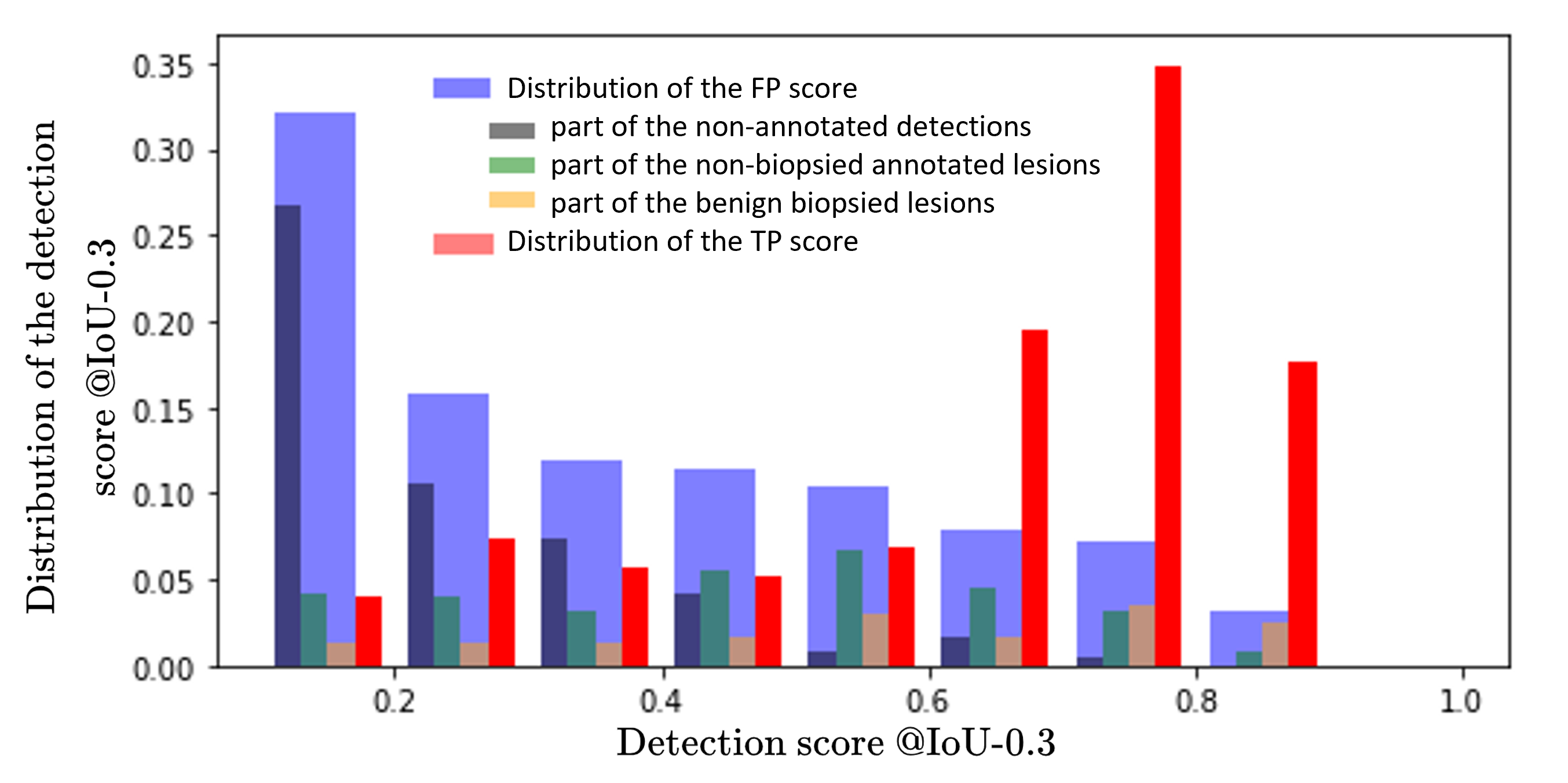}
	\caption{Distribution of the TP (red) and FP (blue) scores (in the range [0.1-1]) for all detections on the test set. The FP are separated into three categories: non-annotated lesions, annotated non-biopsied and annotated biopsied benign lesions. A score of 0.1 corresponds to 95\% detection sensitivity.
	}
	\label{fig:0} 
\end{figure}

For the detection of cancers, the distribution of the model scores for true positive (TP) or false positive (FP) detections are shown figure~\ref{fig:0} with respectively the red and blue bars. 
When a FP is detected, it is interesting to separate (black) the non-annotated detection, (green) the non-biopsied annotated lesions and (orange) the biopsied benign lesions. With a high detection score ($>$0.7), the FP are mainly composed of suspicious lesions that turned out to be benign after biopsy (orange bar). As they were labeled 'suspicious' the radiologist biopsied them and it was expected that those lesions could be confused with the malignant ones.

Figure ~\ref{fig:1}(b) corresponds to the ROC curve for the cancer classification at the breast level and the blue dots are clinical results obtained from various radiologists classifying CESM images\cite{zhu2018diagnostic}. 
%(median classification sensitivity from 18 clinical studies $\bar{\text{Se}}$=0.952). 
The AUC of the deep learning breast classification model is 0.930. The median specificity for the 18 clinical studies is 0.757. At this fixed specificity, the sensibility of the AI model is 0.929.
%\SM{En quoi l'AUC peut elle être comparée avec la Se médiane d'un ensemble de lecteurs? Il serait peut être plus intéressant de donner la Se (avec CI) du modèle pour la Sp médiane des lecteurs.}

\begin{figure}[htbp!]
	\centering
        \includegraphics[width=1\textwidth]{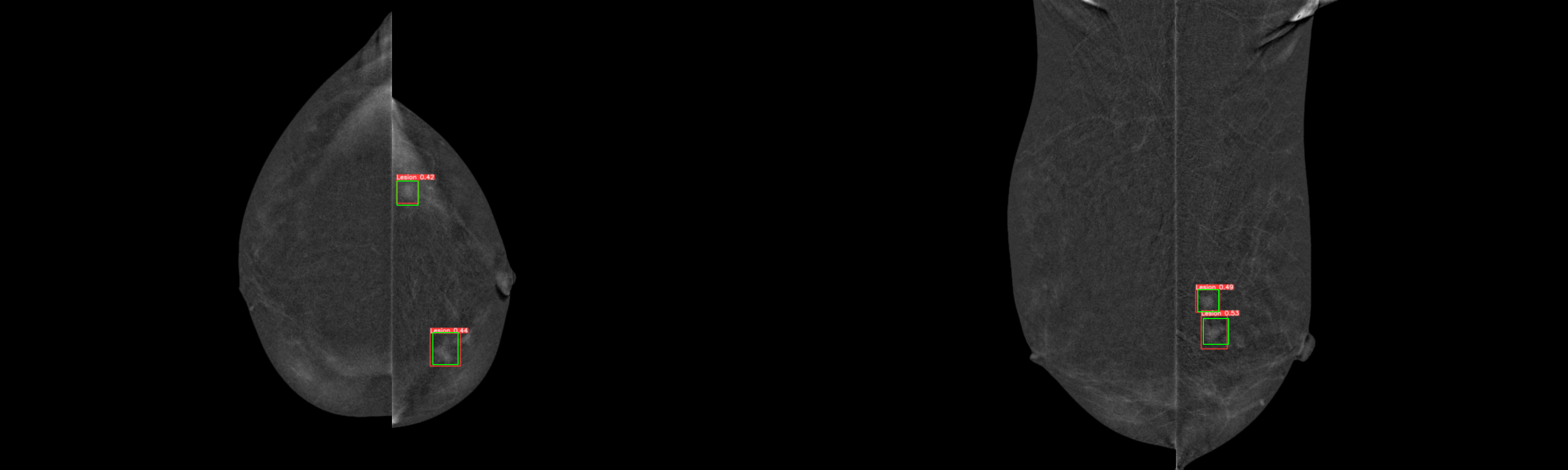}
        \includegraphics[width=1\textwidth]{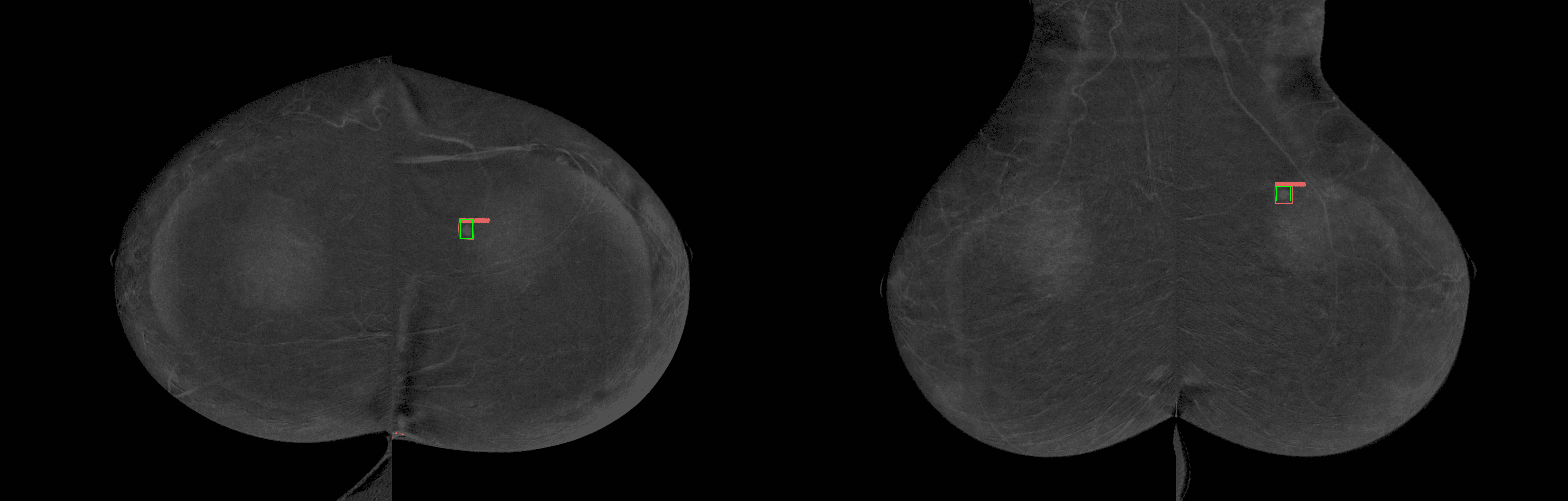}
        \includegraphics[width=1\textwidth]{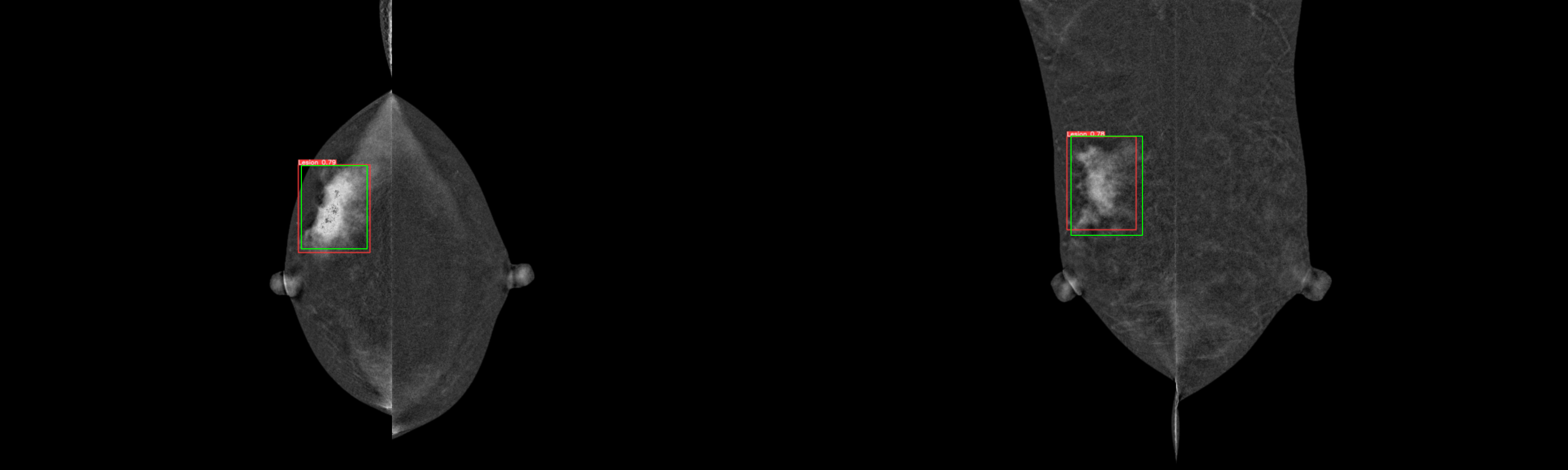}
        \includegraphics[width=1\textwidth]{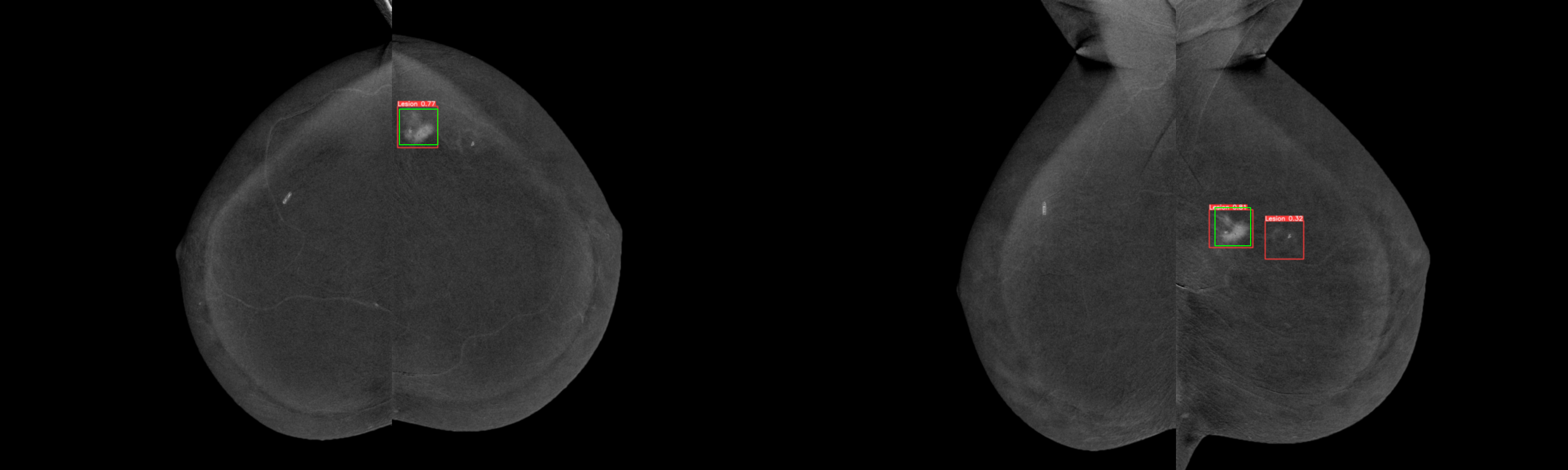}
	\caption{
	Results for lesion detection at Se=0.95. Red box: detected area, green box: ground truth. In addition, a suspicion score is associated to each ROI.}
	\label{fig:2} 
\end{figure}

The lesion detection results obtained for few patients are illustrated in figure~\ref{fig:2} with the red boxes being the detected regions and the green boxes the annotated ground truth. It can be seen that the detections are mainly consistent in the CC and MLO views (similar structure detection when the lesion was visible in the two views). However, the model was designed to detect lesions for each view independently. In the last presented patient for example, a FP on the LMLO view was not seen on the LCC view. Developing a multi-view CAD would be of upmost interest to increase the CAD efficiency. 

\begin{figure}[!htbp]
	\centering
        \includegraphics[width=0.7\textwidth]{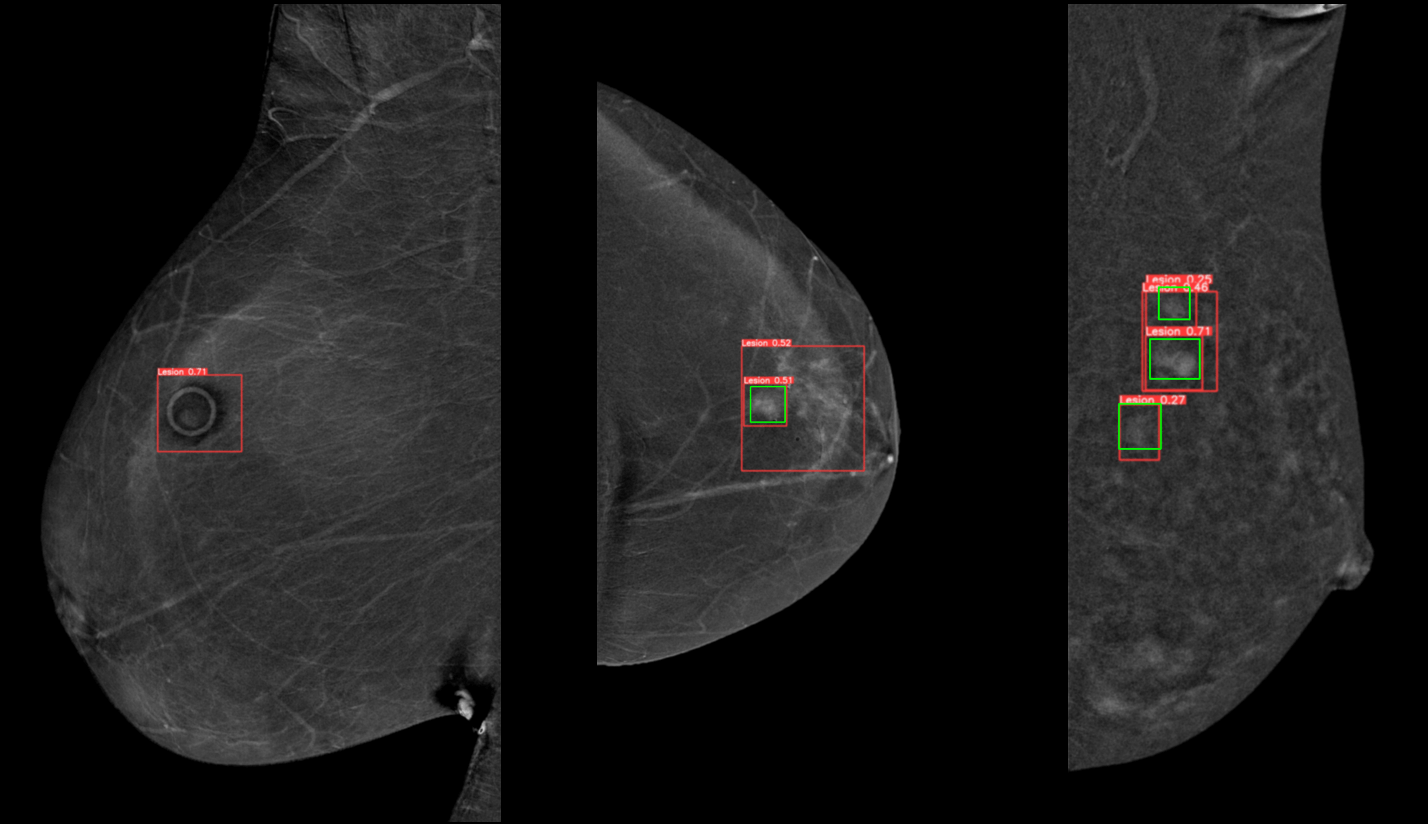}
	\caption{Three images containing FP detections (score threshold for cancer detection Se=0.95). (left) detection of a foreign object, (center) two detections of the contrast uptake, the smallest corresponds to the annotated ground truth while the largest is considered as a FP and (right) the two detected lesions are supplemented with another third large box that is considered as a FP.
	}
	\label{fig:3} 
\end{figure}
Figure~\ref{fig:3} presents 3 views from three patients with FP detections. On those images the ground truth lesions are shown in green and the detected boxes in red. Although the ground truth lesions were correctly identified, multiple false positives are also detected. 
When analyzing the FP detections, three main categories appear: 
\begin{itemize}
    \item detection of localized non-lesion patterns (\eg vessels, biopsy clips, foreign objects, mole markers, \etc). An example of a non-lesion detection is shown figure~\ref{fig:3} (left).
    \item detection of fragmented or complex lesions where the annotation is inconsistent with the detection (\eg a micro-nodular contrast uptake annotated as a single box and detected with multiple ROIs). An example of this FP is shown figure~\ref{fig:3} (center and right).
    In some cases, all fragmented lesions are detected and supplemented by a broad general annotation. 
    These kinds of inconsistent detections may generate a lot of FP.
    \item detection of CESM artifacts (\eg scars, skin folds). For this reason, it is very important to use CESM recombined images with the least artifacts (\eg processed with Nira algorithm).
\end{itemize}

% =========================================================
\section{Conclusion}
% =========================================================

This study presents the development of a deep learning-based detection model in CESM images. 
A large CESM dataset composed of 586 patients (for a total of 2510 images) collected from various hospitals and acquired with different imaging systems from GE Healthcare is used. All labels were biopsy proven. The detection model is a small Yolo-v5 architecture, trained with data augmentation strategies. Various metrics were used to assess the efficiency of the detection and image classification.
The trained model performed breast classification with an AUC of 0.93 and was able to identify cancers with a sensitivity of 0.95 and 0.61 false positive per image. 

One perspective of this work is to include multi-view detection~\cite{yang2021momminet} as out current model detects lesions independently of views. Unifying the detections should improve model performance. 

Considering both the recombined and the low energy images will be a future investigation. As performed in CESM ROI classification papers\cite{massafra2021radiomic}, the low energy may contain discriminant information to classify the detected ROIs.

The model we developed has the potential to be used for triaging patients for biopsy or follow-up, and to help the radiologist detecting complex contrast uptake that could modify the treatment planning (\eg detection of satellite lesions). 
With already promising results, we aim to extend this study to a larger database to continue improving the model efficiency and robustness.

The originality of this study lies in two contributions:
\begin{itemize}
    \item While deep learning models have been a breakthrough for lesion detection and analysis in FFDM and DBT images, they have not been employed in CESM. This work is, to the best of our knowledge, the first attempt to develop a deep-learning CAD for CESM.
    \item Collected from various sites, this CESM dataset with ground truth used in this study is remarkably large compared to the literature. 
\end{itemize}

% =========================================================
\section{Acknowledgements}
% =========================================================
The authors would like to acknowledge Ruben Sanchez, Ann-Katherine Carton, Jean-Paul Antonini (all working at GE Healthcare, Buc, France) for their help in data collection and insights in mammography and Andrei Petrovskii for useful discussion on deep learning models. We wish to point out the absence of conflicts of interest related to our study.

% =========================================================
\section{Compliance with Ethical Standards}
% =========================================================

This research study was conducted retrospectively using anonymized human subject data made available by research partners. Applicable law and standards of ethic have been respected.

\bibliographystyle{spiebib} 
\bibliography{Biblio}

\end{document}